\documentclass[11pt,twoside]{article}

\usepackage{asp2006}
\usepackage{epsf}
\usepackage{psfig}
\usepackage{lscape}

\begin{document}
\title{Structures in galaxies: nature versus
  nurture.\\ Input from theory and simulations}   
\author{E. Athanassoula}   
\affil{LAM/OAMP, UMR6110, CNRS/UP,
Technop\^ole de Marseille Etoile,\\ 38 rue Fr\'ed\'eric Joliot Curie,
13388 Marseille C\'edex 20, France\\}    

\begin{abstract}
Galaxies, in particular disc galaxies, contain a number of
structures and substructures with well defined morphological, photometric
and kinematic properties. Considerable theoretical effort has been put
into explaining  
their formation and evolution, both analytically and with numerical
simulations.  In some theories, structures
form during the natural evolution of the galaxy, i.e. they are a
result of nature. For others, it is the interaction with other
galaxies, or with the intergalactic medium -- i.e. nurture -- that accounts
for a structure. Either way, the existence and properties of these
structures reveal important information on the underlying potential
of the galaxy, i.e. on the amount and distribution of matter
-- including the dark matter -- in
it, and on the evolutionary history of the galaxy. Here, I will briefly
review the various formation scenarios and the respective role of
nature and nurture in the formation, evolution and properties of the
main structures and substructures.   

\end{abstract}

\vspace{-0.5cm}
\section{General introduction}

What do we mean by ``nature versus nurture''? To some degree, all galaxies
have one or more neighbours 
and/or are influenced by their surroundings. So in the
absolute sense of the word, there is no such thing as an isolated
galaxy, except in computer simulations. On the other hand, even though
a given structure may be triggered by an interaction, it will
evolve within its galaxy potential, which, together with 
the remaining characteristic properties of the galaxy, will influence
its evolution. 
Thus, no structure is 100\% due to nature, or 100\% due to nurture. 
It is, nevertheless, useful to ask which of the two has mainly
influenced the formation and evolution of a given structure, and,
therefore, in the following I will assign the origin of a structure to
one of these two alternatives. 

In fact, as will be seen below, more than one 
scenario is possible for each structure, and, very often, 
some mixture of nature and nurture can be involved. Hence, one can
only reason in terms of probabilities and propose this or that agent
as the most probable cause for a given feature. Time comes into play
as well: if one goes sufficiently far back, all galaxies have
experienced important interactions with their surroundings. If,
however, the time elapsed since the last interaction is longer than
the time necessary for a given 
structure to form, one can safely argue that this structure is due to
nature rather than to nurture (see also a discussion in
\cite{HopkinsKMQ09b}). 

After these introductory remarks, let me review what models
and simulations can tell us. Since the subject is very broad,  
I will discuss here only certain aspects, necessarily 
reflecting my own preferences, without me wanting to belittle
in any way work which I cannot mention for lack of time and space. 
The list of references given is likewise far from complete. 
Finally, I will only discuss the formation of structures and not their
destruction, although the latter could also be due either to nature or
to nurture. 

\section{Discs}

Formation of galactic discs is one of the major still unsolved
astronomical problems. Very schematically, one can distinguish 
two families of theories, both with important
implications for the evolution and development of structures and
stellar populations in disc galaxies.

The first one explains disc formation as due to 
gas inflow, which can come either in the form of
a gradual accretion, or of minor gas\--rich mergers
\citep[][etc.]{White.Rees.78, Fall.Efstathiou.80, MoMW98, 
  Dekel.Birnboim06, HellerSA07, Genzel08,
  AgertzTM09, Bournaud.Elmegreen.09, Epinat09, Law09}. 

More recently, however, a second alternative has been proposed, which, 
if confirmed, would make discs the result of major mergers.
Since the pioneering work of \cite{Toomre.Toomre.72} and 
\cite{Toomre.77}, many arguments, mainly based on simulations, have
shown that a dry major merger 
will result in an elliptical galaxy. The outcome of wet major mergers 
has been relatively less well studied and is more complex, since 
it can result in a disc plus spheroid. The mass ratio of these two
components depends not
only on the gas fraction in the progenitors and their mass ratio, but
also on the encounter orbit, the relative orientation of their discs
and even on the properties of their gaseous component (particularly its
feedback). A merging of present day spirals, even late types, will fall
short of the necessary gaseous content \citep{Barnes02}.  
At higher $z$, however, progenitors are much more gas rich, and the final
merger can be a disc
galaxy \citep[e.g.][]{Springel.Hernquist.05,
  Robertson06, Governato08, Hopkins09a, StewartBWM09}.  

More work is needed before the formation of discs by mergers can be
considered as established. Nevertheless, it is interesting to point
out that this second formation scenario, which has considerable
support from observations \citep{HammerFEZLC05,
HammerFPARYD09}, would result in younger discs. Furthermore,
present day discs, and therefore their structures and substructures,
would then be of a second generation following the ones formed
initially in their progenitors. 

\section{Bulges}

Bulges are a very inhomogeneous class of objects
\citep{Kormendy.Kennicutt04}. \cite{Atha05a} reviews the various
definitions of bulges and distinguishes three 
types. 
{\bf Classical} bulges form mainly by mergings, have a
spheroidal shape and a very centrally concentrated mass
distribution.
 {\bf Boxy/Peanut} bulges are parts of bars that stick
well out of the disc equatorial plane. They form from vertical instabilities
of the bar material. Finally, {\bf discy} bulges have the
shape of a disc, but are not necessarily axisymmetric. They are due to
disc material that has been pushed inwards either by interactions/mergers,
or by the torques due to a bar. Thus boxy/peanut bulges can be
primarily assigned to nature, classical bulges primarily to nurture and discy
bulges to both. 

\section{Bars}

$N$-body simulations have shown that stellar bars form naturally in galactic
discs \citep[for recent reviews see][]{Atha05b,
  Binney.Tremaine.08} and that 
their growth rate depends on the halo-to-disc mass 
ratio -- measured in the inner parts of the galaxy -- and the velocity
dispersion in the disc \citep{Atha.Sellwood.86}. Thus, stellar bars grow
faster in relatively more 
massive and colder discs. They then evolve by redistributing the
angular momentum within the galaxy. This is emitted mainly from 
near-resonant material in the bar region and absorbed mainly by near-resonant 
material in the halo and the outer disc, so that the bar
strength correlates well with the amount of angular momentum exchanged
\citep{Atha02,Atha03}. 
The pattern speed of such bars decreases with time and its 
evolution is followed by a number of morphological,
kinematic and photometric changes, the most spectacular of which is
the formation of
a boxy/peanut bulge from the vertical instability of the bar
\citep{CombesDFP90, RahaSJK91, Debattista.Sellwood.00, Atha.Misiriotis02,
  Valenzuela.Klypin03, ONeill.Dubinski.03, Atha03, Atha05a, DebattistaMCMW06,
  MartinezVSH06}. For the formation and evolution of bars in gas rich
systems see e.g. \cite{BournaudCS05}, \cite{BerentzenSMVH07},
\cite{RomanoDSHH08} and references therein.

Thus, bar formation and evolution can be explained as
totally due to nature. Nevertheless, 
interactions also can trigger bars, or can strongly influence bar
properties \citep[e.g.][]{GerinCA90,
  Miwa.Noguchi98, BerentzenAHF03, BerentzenAHF04}. For example,
impacts by small companions 
can produce off-centred bars, which are generally shorter and less
elongated than their centred precursors \citep{AthaPB97}.  

\section{Spirals}

In cases like M51, it is clear that the spiral has been triggered by
an interaction 
\citep{Toomre.Toomre.72,Toomre81}. This, however,
is not true 
in all cases and several mechanisms have been proposed to form
spirals in isolation. 

The SWING mechanism \citep{Toomre81} relies on the amplification of a spiral as
it swings from leading to trailing in the corotation region of a
shearing galactic disc. It involves an outwards travelling
and swinging leading wave and two trailing waves, one travelling
outwards from corotation to the outer Lindblad resonance and the other
inwards. If the latter is reflected at the centre of the galaxy or at
a sharp edge, the cycle can close. Thus, this mechanism could
explain both grand design and flocculent spirals. 

In the case of barred galaxies the spirals can also be due
to material guided by the manifolds emanating from the unstable
Lagrangian points at the ends of the bar \citep{RomeroGAMG07,
  AthaRGM09a, AthaRGBM09b}. This theory explains why grand design
spirals in barred 
galaxies are trailing and have preponderantly two arms. It also
reproduces well the shape of the arms.   

Thus, although interactions can link spirals to nurture, other mechanisms 
explain them a result of nature.

\section{Rings}

Like bulges, rings are also an inhomogeneous class of objects
\citep{Atha.Bosma.85}. A small companion, hitting a disc in a
direction not too far from perpendicular to its equatorial plane and
not too far from its centre, will create a  
density wave in the form of a ring, propagating outwards
\citep{Lynds.Toomre.76}. These rings are not common, in agreement with
the fact that they require particular impact parameters. 
Often a second ring is visible. The most spectacular case is the
Cartwheel galaxy, which, besides the first and second rings, has
also spokes between the two. Such rings can be clearly attributed to
nurture. 
 
A different type of rings, not due to a collisional encounter, is 
often observed in barred galaxies. Depending on their location within
the galaxy, these can be classified as nuclear, when they are in the
innermost parts of the disc, inner rings, which have roughly the size of
the bar, and outer ones, which have roughly twice that size
\citep{Buta95}. The inner and outer rings 
can be due to material guided by the manifolds emanating
from the unstable Lagrangian points at the ends of the bar
\citep{RomeroGMAG06, RomeroGAMG07, AthaRGM09a, AthaRGBM09b}, i.e. 
they can
be explained by the same theory that explains spiral arms in such
galaxies. This theory predicts the correct sizes and orientations for
both inner and outer rings. In particular, inner rings are
elongated along the bar and outer ones either along it or
perpendicular to it, as observed. This theory also explains the rings
found in gas response simulations as those of Schwarz, Salo, Byrd,
Laurikainen, Rautiainen, etc., whose results agree very well with
observations. 
 
Polar rings are yet a third type of rings. They are external to the
disc with a size of the order of twice the disc radius and their plane
makes usually an 
angle of roughly 90$^{\circ}$ with that of the disc. For this
reason, they are generally believed to be due to the accretion of a
companion galaxy, or to mass transfer during encounters, i.e. due to
nurture. 

\section{Other features}

\subsection{Thick discs}

Disc galaxies often have a second disc, aligned with the standard,
well-known thin disc, with the same equatorial plane, but much thicker
\citep[][etc.]{Tsikoudi79, Burstein79}. Its existence has been,
directly or indirectly, attributed 
to mergings with small companions, or satellites captured by the
disc. The material that constitutes it 
would then be either from the pre-existing thin disc, which has been
puffed up by the merging, or directly from the companion, which is
disrupted and adopts the orientation of the initial thin disc, or
from both \citep[e.g.][]{WalkerMH96,
  PenarrubiaMB06}. Either way, thick discs are attributed to 
nurture. 

\subsection{Shells}

Shells around an elliptical galaxy can be well explained as due to the
disruption of a small companion falling in the elliptical
\citep[see][for a review]{Atha.Bosma.85}. Similarly, small central
discs in such galaxies can be due to gas-rich infalling companions.

\subsection{Warps}

Although several attempts have been made to explain warps as due to
nature, I will here focus on the nurture origin, which is more
generally followed. Two such possibilities have been mainly
considered. The first possibility \citep{Ostriker.Binney.89, Quinn.Binney.92,
  Debattista.Sellwood.99, Jiang.Binney.99, Shen.Sellwood.06} is that warps
are due to cosmic infall. If this reorientates the outer parts of the
halo by several degrees per Gyr, the disc will develop a warp with
properties similar to those observed. 

The second and most straightforward way of producing warps is through
the tidal interaction with a close companion, as has been witnessed in
a large number of simulations. If this alternative is generally
true, i.e. if this is the way that all warps are formed, then we
should be able to observe around all warped discs a companion whose
mass and orbit can account for the warp. NGC 4013 was for a 
long time believed to be a counterexample to 
this possibility. Yet deep imaging in \cite{MartinezPGMPP09} 
showed the debris of a past companion, so 
that even in this case the warp could be tidally generated.

\subsection{Bridges and tails}

Spectacular examples of such structures can be seen in M51, or the
antennae. 
After the pioneering work of \citet{Toomre.Toomre.72} and the
many other studies that followed it, it has
become clear that these structures are due to gravitational
interactions, i.e. are due to nurture. 

\section{Summary}

I discussed the possible formation mechanisms of structures and
substructures in galaxies, focusing on whether their origin is due to
nature or to nurture. In most cases, there is more than one alternative.
A few structures can be explained only by a mechanism relying on
nurture, but none uniquely only to nature.
 
\acknowledgements 
I thank the organisers for inviting me to to this very interesting and
stimulating meeting. This work was partly supported by grant
ANR-06-BLAN-0172. 

%%% THE BIBLIOGRAPHY
%%%
%%% CONSULT SECTION 3 OF "INSTRUCTIONS FOR AUTHORS" FOR HOW TO USE NATBIB.
%%% AUTHORS ARE ENCOURAGED TO USE EITHER THE "THEBIBLIOGRAPY" ENVIRONMENT
%%% BY UNCOMMENTING (DELETING THE "%" SYMBOL) THE COMMANDS BELOW, OR BY
%%% USING THE BIBTEX ENVIRONMENT. TO FIND OUT WHICH IS APPLICABLE TO YOUR
%%% CONTRIBUTION, CONSULT THE VOLUME EDITORS FOR YOUR PROCEEDINGS.
%%%

\end{document}